\documentclass[conference]{IEEEtran}

\ifCLASSOPTIONcompsoc
  \usepackage[nocompress]{cite}
\else
  \usepackage{cite}
\fi
\usepackage[T1]{fontenc}
\usepackage{changepage}
\usepackage{adjustbox}
\usepackage{tablefootnote}
\usepackage[utf8]{inputenc}
\usepackage{mathrsfs}
\usepackage{stmaryrd} 
\usepackage{tabularx,lipsum,environ,amsmath,amssymb}
\usepackage[boxruled,linesnumbered]{algorithm2e}

\makeatother
\usepackage{hyperref}
\usepackage{cleveref}
\crefname{Lemma}{Lemma}{Lemmas}
\usepackage{booktabs}
\usepackage{algorithm2e}
\usepackage[table,xcdraw]{xcolor}
\usepackage{tabularx,lipsum,environ,amsmath,amssymb}
\usepackage{caption}
\usepackage{algorithm2e}
\usepackage{graphicx}            
\usepackage{amsmath}             
\usepackage{amssymb}
\usepackage{array}
\usepackage{layout}
\usepackage[margin=1.04in]{geometry}
\usepackage{amsmath,amssymb}
\usepackage{gensymb}

\makeatletter
\newsavebox\myboxA
\newsavebox\myboxB
\newlength\mylenA

\newcommand*\xoverline[2][0.75]{%
	\sbox{\myboxA}{$\m@th#2$}%
	\setbox\myboxB\null
	\ht\myboxB=\ht\myboxA%
	\dp\myboxB=\dp\myboxA%
	\wd\myboxB=#1\wd\myboxA
	\sbox\myboxB{$\m@th\overline{\copy\myboxB}$}
	\setlength\mylenA{\the\wd\myboxA}
	\addtolength\mylenA{-\the\wd\myboxB}%
	\ifdim\wd\myboxB<\wd\myboxA%
	\rlap{\hskip 0.5\mylenA\usebox\myboxB}{\usebox\myboxA}%
	\else
	\hskip -0.5\mylenA\rlap{\usebox\myboxA}{\hskip 0.5\mylenA\usebox\myboxB}%
	\fi}
\makeatother

\usepackage{etoolbox}

\makeatletter
\patchcmd{\@maketitle}{\raggedright}{\centering}{}{}
\patchcmd{\@maketitle}{\raggedright}{\centering}{}{}
\makeatother

\newcommand{\twodots}{\mathinner {\ldotp \ldotp}}

\newfont{\bbb}{msbm10 scaled 700}

\newfont{\bb}{msbm10 scaled 1100}
\newcommand{\CC}{\mbox{\bb C}}

\newcommand{\av}{{\bf a}}
\newcommand{\bv}{{\bf b}}

\newcommand{\uv}{{\bf u}}
\newcommand{\wv}{{\bf w}}
\newcommand{\vv}{{\bf v}}

\newcommand{\Dm}{{\bf D}}

\newcommand{\Sm}{{\bf S}}
\newcommand{\Tm}{{\bf T}}
\newcommand{\Um}{{\bf U}}

\newcommand{\Vm}{{\bf V}}

\newcommand{\herm}{{\sf H}}

\begin{document}
\captionsetup[figure]{labelfont={},labelformat={default},labelsep=period,name={Fig.}}
\title{Power Transfer between Two Antenna Arrays in the Near Field}

\author{\IEEEauthorblockN{Krishan K. Tiwari, Giuseppe Caire}
	\IEEEauthorblockA{Technische Universit\"at Berlin, 10587 Berlin, Germany}
	Email addresses: {lastname}@tu-berlin.de
}
\maketitle

\begin{abstract}
We present numerical results with a focus on power transfer between two standard linear antenna arrays placed in the near field, where a much smaller active multi-antenna feeder (AMAF) space feeds a far larger passive array referred to as a reflective intelligent surface (RIS). The interest is in the regime of focal length to diameter ratio ($F/D$) less than unity. We address the question of center feed vs. end feed for array fed array antenna architectures and present the following novel findings and contributions: 1. In the regime of $F/D$ ratio less than one, the AMAF-RIS power transfer deviates from the classical inverse square law. Furthermore, the behavior of the power transmission coefficient is more sensitive to the $F/D$ ratio than to a particular RIS size. 2. For an end feed, non-eigenmodes provide better beam shapes than the eigenmodes, which are still inferior to beam shapes from the center feed eigenmodes. 3. The center feed provides more power gain than an end feed. This clearly illustrates that the center feed configuration should be used for array fed arrays, in contrast to the classical parabolic geometry. 
\end{abstract}


\section{Introduction}
\label{sec:intro}

\begin{figure}[ht!]
\centerline{\includegraphics[width=7.95cm]{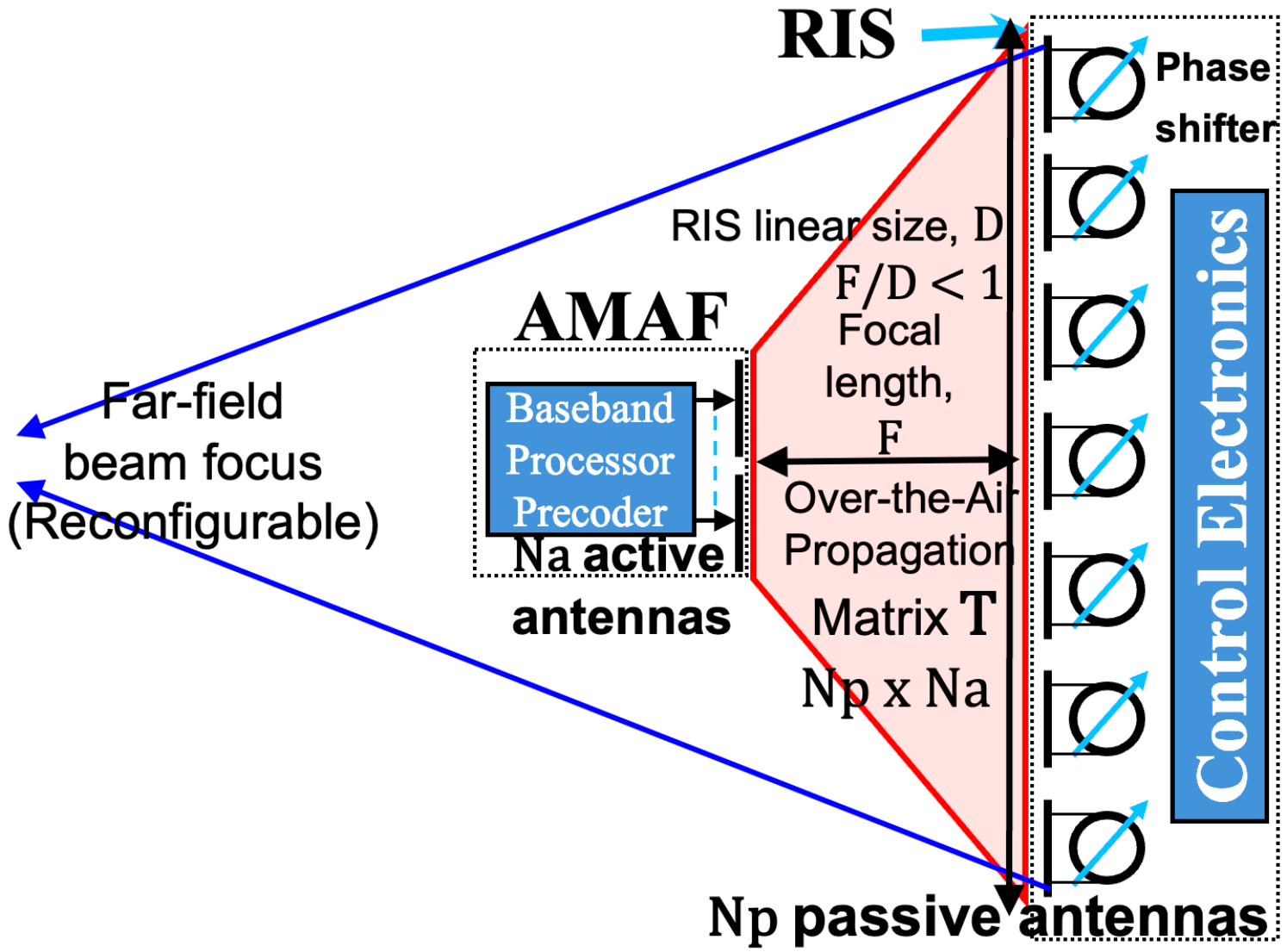}}
	\caption{An Active Multi-Antenna Feeder (AMAF) center feeding a RIS\cite{ICC2023}, in the $F/D<1$ regime. AMAF and RIS are $N_a$ and $N_p$ elements respectively, standard linear arrays\cite[p. 51]{trees} in a 2D geometry for simplicity and better intuition.}
	\label{fig:sysmo}
\end{figure}

\IEEEPARstart{O}{riginally} proposed in the 1960s \cite{60s}, a reconfigurable reflectarray (also known as a ``reflective intelligent surface'' (RIS)) offers the best of two worlds -- the natural hardware efficiency of space feeding (which in turn translates into smaller size, weight, power, and cost (SWaP-C),\footnote{Recall that space feeding has obvious advantages of hardware simplicity and lower SWaP-C \cite{pozar97}, \cite[Table 1]{SWaP-C} than guided wave feeding as in classical constrained-fed large active arrays, e.g., phased arrays. This is particularly relevant in mmWave and sub-THz bands, where RF power levels are limited, interconnect losses are typically high, circuit designs are challenging, and channels are sparse in beamspace (typically the Line-of-Sight).} and higher reliability (due to fewer failure modes, much easier thermal management, lower wind load for flat profiles, etc.)) and the electronic steerability of classical phased arrays \cite{pozar97}. For legacy reasons (starting from (before) Berry's 1963 reflectarray when microstrip patch antenna hardware was not realized), the RIS is fed by a horn antenna. It is prudent to replace the feed horn by a low profile, smaller, light weight planar array antenna, e.g., as in \cite{Webinar, TICRA, TICRA1}. However, all the elements of the feeder array in \cite{Webinar, TICRA, TICRA1} are fed by the same amplitude and the same phase. We have shown in \cite{ICC2023} that configuring the feeder array along the eigenmodes of the AMAF-RIS propagation matrix $\Tm$ (Fig. \ref{fig:sysmo}) provides better and more agile beam shaping capabilities\footnote{Note that replacing the feed horn with the AMAF provides more degrees of freedom by discretizing the feed aperture.}, in addition to the fact that the principal eigenmode results in maximum power transfer.


The classical offset fed parabolic reflector antenna allows a lower profile than its center fed counterpart without compromising the power transfer from the feed horn to the reflector, thanks to the parabolic geometry. Perhaps along similar lines, end feed has been used for reflectarrays fed by an active feeder array, e.g., in \cite{TICRA, TICRA1}. In this paper, we address the question of center feed vs. end feed for array-fed array antenna architectures and present the following novel findings and contributions: 1. In the $F/D<1$ regime, AMAF-RIS power transfer  $\left(\sigma_1^2~ \text{and}~ \sum_{i=1}^{N_a} \sigma_i^2,~ \text{$\sigma_i$ is the $i^{\rm th}$ singular value of $\Tm$}\right)$ deviate from the classical inverse square law and are smaller due to the taper. 2. For an end-feed, non-eigenmodes provide better beam shapes than the eigenmodes, which are still inferior to beam shapes from the center-feed eigenmodes. 3. The center feed provides more power gain than an end feed.



\section{System Model} \label{Models}

The presented numerical results are based on the following models. The AMAF-RIS matrix $\Tm \in \CC^{N_p \times N_a}$ is given by
\begin{equation}
\label{eq:T}
\centering
T_{n,m} = \frac{\sqrt{E_A(\theta_{n,m})E_R(\phi_{n,m}})~ \text{exp}\big(j\pi r_{n,m}\big)}{2\pi r_{n,m}},
\end{equation}
where $E_A, ~E_R, ~\theta_{n,m},~ \phi_{n,m}, ~ \text{and}~r_{n,m}$ are AMAF element power radiation pattern, RIS element power radiation pattern, angle of departure from an AMAF element $m$ to a RIS element $n$ measured from the AMAF element boresight, angle of arrival from the AMAF element $m$ to the RIS element $n$ measured from the RIS element boresight, and the distance between the AMAF element $m$ and the RIS element $n$, respectively, and $j=\sqrt{-1}$. Note that (\ref{eq:T}) is the Friis transmission equation \cite[Eq. (2-119)]{Balanis_antenna_theo} in the magnitude form along with the inclusion of the distance dependent phase term. Note also that, while the AMAF is in the near field of the RIS, the individual antenna elements of the AMAF and the RIS are in the far field of each other. 

The AMAF can be configured along ``eigenmodes'' based on the singular value decomposition (SVD) of $\Tm=\Um\Sm\Vm^\herm$ where $\Um \in \CC^{N_p \times N_p}$ and $\Vm \in \CC^{N_a \times N_a}$ are unitary matrices and $\Sm \in \CC^{N_p \times N_a}$ is a diagonal matrix containing ordered singular values $\sigma_1, \sigma_2, \twodots, \sigma_{N_a}$. $\uv_i$ and $\vv_j$ are the $i^{\text{th}}$ and the $j^{\text{th}}$ column vectors of $\Um$ and $\Vm$, respectively. The principal eigenmode configuration of the AMAF with $\vv_1$ results in maximum power transfer. Note that because $\vv_j$ and $\uv_i$ are unit norm vectors, if a power $P_T$ is provided at the AMAF configured along $\vv_1$, the RIS will be fed by the power $\sigma_1^2 P_T$. For simplicity, we assume $P_T = 1$. Then the principal eigenmode will give the RIS power of $\sigma_1^2$, the second eigenmode will give the RIS power of $\sigma_2^2$, and so on and so forth. Notice that this allows an elegant and clean \emph{single coefficient} representation of the net power transfer between the AMAF and the RIS, unlike the classical horn-fed reflectarray where the analysis becomes complex as it depends on the spillover efficiency, illumination efficiency, etc., \cite[Section 4.4]{ref_wiley_2007}, \cite[Section III, Eqs. (18), (19)]{pozar97}. 

Thanks to \cite{Comp}, array theory based results are a good and reliable match to full-wave solver based results for array antennas (unless very fine accuracies in far-off sidelobes and cross-polarization values are required, which is not the focus of this paper). Therefore, we'll use array theory and pattern multiplication based power radiation pattern plots given by $\big|\av^\herm(\vartheta)\wv\big|^2 E(\vartheta)$, where $\vartheta$ is the angle from the array broadside, $\wv$ is the array beamforming vector, and $\av(\vartheta)=\left[1,e^{j \pi \text{sin}(\vartheta)},\twodots,e^{j \pi (N_p-1)\text{sin}(\vartheta)}\right]^\herm$ is the array steering vector, and $[\cdot]^\herm$ denotes the conjugate transpose. The RIS and the AMAF arrays consist of microstrip patch antenna elements modeled by the classical axisymmetric power radiation pattern, \cite[Eq. (2-31)]{Balanis_antenna_theo}, \cite[Eq. (17)]{pozar97}, $E({\vartheta})=4~\text{cos}^2(\vartheta)$, where $\vartheta$ is also the angle from the element boresight, the half-power beamwidth (HPBW) is 90$^\circ$, and the element gain is 6 dBi.

\section{Numerical Results and Discussions}

Numerical experiments were performed for different values of $N_p$ and the focal-length-to-diameter ($F/D$) ratio. Focal length ``$F$'' is the AMAF-RIS distance and ``$D$'' is the RIS array size, as in Fig. \ref{fig:sysmo}. All lengths/distances are normalized to half wavelength. We understand the AMAF-RIS power transfer for the center feed first, in view of its promising beamshaping capability using different eigenmodes as illustrated in \cite{ICC2023}\footnote{Recall from \cite[Section IV]{ICC2023} that the AMAF size $N_a=4$ provides very low side-lobe level ($<$ -50 dB) far-field patterns as in \cite[Fig. 3]{ICC2023} and also allows four eigenmodes of $\Tm$, enabling multiple functionalities, e.g., monopulse-like angular tracking, flat-top beams, etc. Therefore, we keep $N_a=4$.}, in subsection \ref{subsec:cf}, and then take up the end feed in subsection \ref{subsec:ef}.

\subsection{Center Feed}
\label{subsec:cf}
\begin{table*}[ht]
\renewcommand*{\arraystretch}{1.5}
\caption{Center feed: $\sigma_i^2$ in the $F/D < 1$ regime.}
  \begin{adjustwidth}{-1in}{-1in}
    \centering
    \begin{adjustbox}{width=0.9\textwidth}
\label{tab:sigma_f}
\centering
\begin{tabular}{ccccccccccc}

\hline\hline
\text{Sl. No.} & \textbf{$N_a$} & \textbf{$N_p$} & $F$ & \textbf{$\sigma_1^2$} (dB) & \textbf{$\sigma_2^2$} (dB) & \textbf{$\sigma_3^2$} (dB) & \textbf{$\sigma_4^2$} (dB) & \textbf{$\sum_{i=1}^4 \sigma_i^2$} (dB) & \textbf{$\sigma_1/\sigma_4$} & \textbf{$\sigma_1^2/\sigma_4^2$} (dB)\\
\hline
1  & 4 & 8  & 4   & -7.32  & -8.28  & -10.91 & -18.31  & -3.67     & 3.54   & 10.98  \\
2  & 4 & 8  & 8   & -10.34 & -12.53 & -19.73 & -34.98  & -7.98     & 15.92  &  24.04      \\
3  & 4 & 8  & 40  & -21.14 & -35.13 & -58.03 & -87.63  & -20.97    & 2113.2 &  66.50       \\
4  & 4 & 8  & 80  & -26.99 & -46.95 & -76.00 & -111.64 & -26.99    & 17079.0  &  84.65    \\
5  & 4 & 8  & 120 & -30.48 & -53.96 & -86.55 & -125.71 & -30.46    & 57756  & 95.23 \\
6  & 4 & 16 & 8   & -10.26 & -11.21 & -13.78 & -21.32  & -6.59     & 3.57   & 11.05 \\
7  & 4 & 16 & 16  & -13.31 & -15.42 & -22.48 & -36.96  & -10.90    & 15.22  & 23.65 \\
8  & 4 & 16 & 40  & -18.71 & -26.84 & -43.10 & -66.18  & -18.07    & 236.27 & 47.47 \\
9 & 4 & 16 & 80  & -24.14 & -38.08 & -60.77 & -89.90   & -23.98    & 1940.1 & 65.76\\
10 & 4 & 16 & 120 & -27.54 & -44.98 & -71.27 & -103.92 & -27.45    & 6587.2 & 76.37\\ 
11 & 4 & 32 & 8   & -10.25 & -11.17 & -13.08 & -17.69  & -6.25     & 2.36   &  7.46\\
12 & 4 & 32 & 16  & -13.25 & -14.2  & -16.76 & -24.33  & -9.58     & 3.58   & 11.08\\
13 & 4 & 32 & 32  & -16.31 & -18.4  & -25.43 & -39.87  & -13.89    & 15.07  &  23.56\\
14 & 4 & 32 & 40  & -17.4  & -20.57 & -29.80  & -46.48 & -15.53    & 28.42  & 29.07\\
15 & 4 & 32 & 80  & -21.72 & -29.83 & -46.05 & -69.03  & -21.08    & 231.98 & 47.31\\
16 & 4 & 32 & 120 & -24.81 & -36.29 & -56.32 & -82.83  & -24.56    & 796.34 & 58.02  \\   \hline
\end{tabular}
 \end{adjustbox}
  \end{adjustwidth}
\end{table*}

The numerical experimental results are summarized in Table \ref{tab:sigma_f}. We see in Table \ref{tab:sigma_f} that $\sigma_1^2$ is not the direct isotropic propagation loss at smaller $F/D$ ratios (around 1 and smaller), e.g., for $F/D<1$, when the AMAF-RIS distance $F$ is halved, $\sigma_1^2$ increases by only 3 dB. Recall that the AMAF and the RIS phase centers are their respective centroids \cite{SLAcentroid}. In the regime of $F/D<1$, the end RIS elements do not receive increased power due to a smaller ``$F$'' because the angular spread from the center to the ends increases with decreasing $F$, resulting in a much lower power coupling from the AMAF to the end elements, and therefore when $F$ is halved, $\sigma_1^2$ increases by only 3 dB instead of the expected 6 dB increase due to the reduced isotropic spreading loss at a half distance. This is the reason why the $\sigma_1^2$ behavior, for $F/D<1$, deviates from the inverse square law. In contrast, if $F/D >> 1$, then the 6 dB increase and the inverse square law with focal length ``$F$'' holds in general, because then the angular spread from RIS center to RIS ends is very small and the two arrays see each other more like point sources in the far field.  

\begin{figure}[ht!]
\centerline{\includegraphics[width=8.25cm]{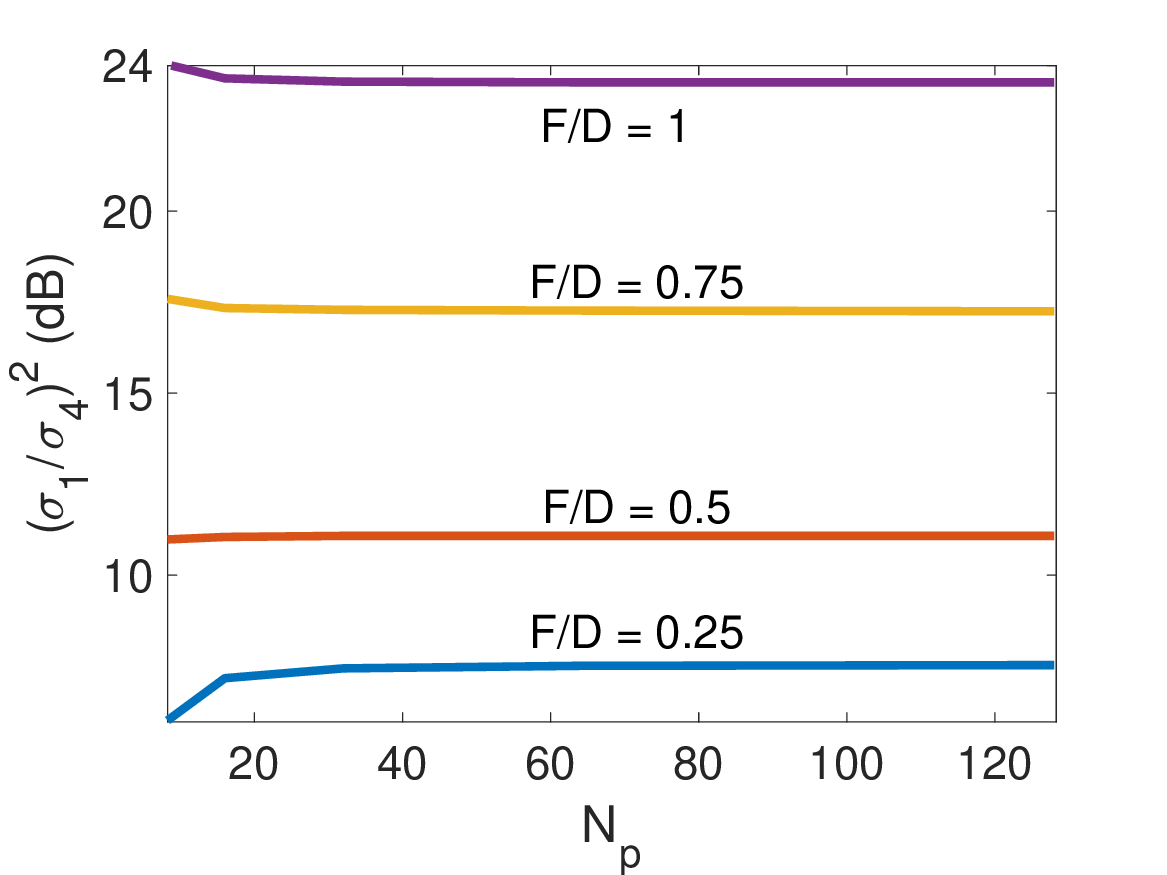}}
	\caption{$\sigma_1^2/\sigma_4^2$ (dB) for different $F/D$ ratios and RIS sizes $N_p$.}
	\label{fig:Bounds_Plot_VTC2024}
\end{figure}

Note also that as the $F/D$ ratio decreases from 1 to 0.5, $\sigma_1^2$ consistently increases by almost 3 dB for $N_p=8, 16, \text{and}~ 32$, which means that the maximum AMAF-RIS power transfer depends on the ``$F/D$'' ratio and not just on the RIS size D per se. Furthermore, we also see in Table \ref{tab:sigma_f} that for a given $F/D$, the condition number of $\Tm$, given by $\sigma_1/\sigma_4$, remains essentially the same for different RIS sizes, e.g., for $F/D = 0.5$, the condition numbers for $N_p = 8, 16, \text{and}~ 32$ are $3.54, 3.57, \text{and}~ 3.58$, respectively. Recall that the maximum and the minimum AMAF-RIS power transfers are given by $\sigma_1^2$ and $\sigma_{Na}^2$, respectively. Thus, the ratio of the upper and the lower bounds of the AMAF-RIS power transfer is given by the square of the condition number for a given $\Tm$. We see in Fig. \ref{fig:Bounds_Plot_VTC2024} that this ratio of the upper and the lower bounds of the power transfer is nearly flat with $N_p$ for a given $F/D$ ratio. 

Similarly, $\big(\sum_{i=1}^{N_a} \sigma_i^2\big) -\sigma_1^2$ is also nearly constant for a given $F/D$, e.g, for $F/D = 0.5$, $\big(\sum_{i=1}^{N_a} \sigma_i^2\big) -\sigma_1^2$ are $3.65, 3.67, \text{and}~ 3.67$ for $N_p = 8, 16, \text{and}~ 32$, respectively. Notice also that as $F/D$ increases from 0.5 to 1, $\sum_{i=1}^{N_a} \sigma_i^2$ increases by 4.31 dB for $N_p = 8, 16, \text{and}~ 32$. Thus, in the $F/D$ regime of interest, i.e., $0.5 < F/D < 1$, the condition number of $\Tm$ and the behavior of the singular values of $\Tm$ are determined by the $F/D$ ratio (instead of a specific RIS size $N_p$). 

However, if $F$ is increased from 40 to 80 for $N_p= 8, 16, \text{and}~ 32$, then $\sigma_1^2$ decreases by 5.85, 5.43, and 4.32 dB, respectively, because the $F/D$ scales are different and decreasing (while $F$ doubles in all the three cases). The $\sigma_1^2$ decrease with the doubling of $F$ is now closer to 6 dB as compared to the regime of interest, i.e., $0.5 < F/D < 1$.

Recall that for an unnormalized distance $d$, the isotropic propagation loss is given by $L_{\rm iso}= ( 4 \pi d / \lambda)^2$, where $\lambda$ is the carrier wavelength. Since all distances in this paper are normalized to $\lambda/2$, the unnormalized AMAF-RIS distance is given as $F\lambda/2$, where $F$ is the AMAF-RIS distance normalized to $\lambda/2$, i.e., $d=F\lambda/2$. This gives us $L_{\rm iso}= ( 2 \pi F)^2$. The isotropic channel coefficients for $F = 4, 8, 16, \text{and}~ 32$ are -28 dB, -34.03 dB, -40.05 dB, and -46.07 dB, respectively. We see from Table \ref{tab:sigma_f} that at $F = 8$, $\sigma_1^2=-10.3$ dB for $N_p = 8, 16, \text{and}~ 32$. The numerical calculation for $N_a=N_p=1$ and $F = 8$ yields $\sigma_1^2=-22$ dB, while $L_{\rm iso}=-34$ dB. This is due to the 6 dBi patch element gain at both the AMAF and the RIS. For $N_a=N_p=2$ and $F = 8$, $\sigma_1^2=-16.51$ dB and $\sum_{i=1}^{N_a} \sigma_i^2=-16.06$ dB where the AMAF array and the RIS array each contribute 3 dB gain. For $N_a=N_p=4$ and $F=8$, $\sigma_1^2=-11.26$ dB, $\sigma_2^2=-17.96$ dB, and $\sum_{i=1}^{N_a} \sigma_i^2=-10.40$ dB. Note that both the AMAF array and the RIS array should ideally contribute 3 dB gain each compared to $N_a=N_p=2$ and $f=8$, for a total gain of 6 dB from $\sum_{i=1}^{N_a} \sigma_i^2=-16. 06$, which should have resulted in $\sum_{i=1}^{N_a} \sigma_i^2=-10.06$ dB. Instead, we get $\sum_{i=1}^{N_a} \sigma_i^2=-10.40$ dB. Again, this is because the end RIS elements are at larger angular offsets from the AMAF broadside, and \emph{the inherent angular selectivity of the directive AMAF and RIS elements prevents the realization of the full array gain} and its manifestation in the $\sum_{i=1}^{N_a} \sigma_i^2$. This effect becomes more pronounced as the RIS size $N_p$ increases to 128. 



Finally, from Table \ref{tab:sigma_f} and further numerical calculations, for $N_a=4, N_p=8$ and $F=8$, $\sum_{i=1}^{N_a} \sigma_i^2 = -7.98$ dB, which is only a 2.42 dB increase higher than $\sum_{i=1}^{N_a} \sigma_i^2 = -10.40$ dB for  $N_a=4, N_p=4$ and $F=8$, as compared to the expected 3 dB increase due to the doubling of $N_p$. For $N_a=4$ and $F=8$, $\sum_{i=1}^{N_a} \sigma_i^2$ converges to -6.59 dB, -6.25 dB, -6.22 dB, and -6.22 dB for $N_p=$ 16, 32, 64, and 128, respectively. Also, $\sigma_1^2$ is -10.34 dB, -10.26 dB, -10.25 dB, -10.25 dB, and -10.25 dB for $N_p=$ 8, 16, 32, 64, and 128, respectively. This demonstrates the impact of increased angular spread in limiting the anticipated increase in $\sum_{i=1}^{N_a}\sigma_i^2$ as $N_p$ increases for a given $F$ and a given $N_a$.

\subsection{End Feed}
\label{subsec:ef}
\begin{figure}[ht!]
\centerline{\includegraphics[width=8.25cm]{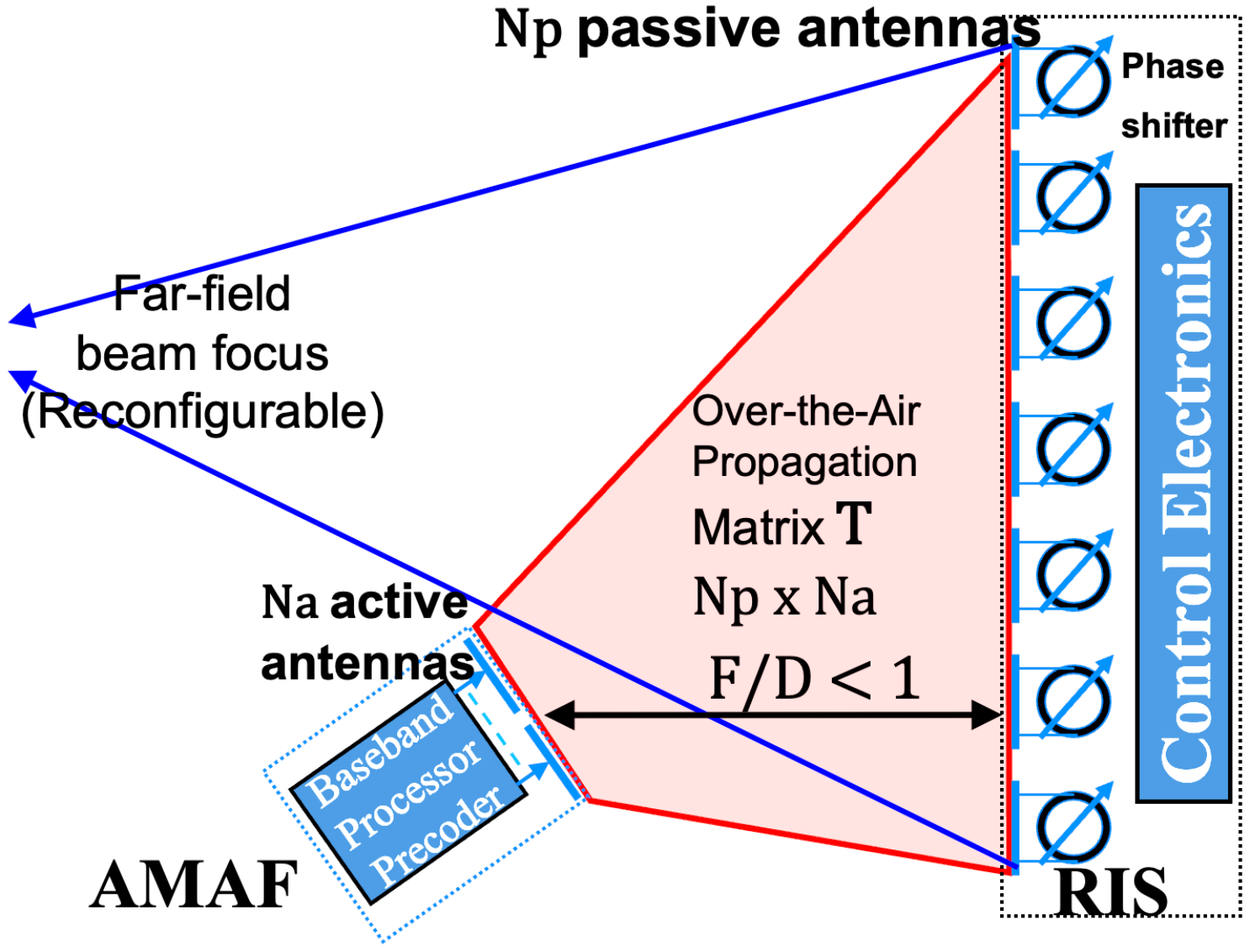}}
	\caption{A tilted AMAF end feeding a RIS.}
	\label{fig:end_fed_sysmo}
\end{figure}

We now consider the end feed as in Fig. \ref{fig:end_fed_sysmo} where the AMAF is tilted such that its mechanical boresight points to the RIS center for a better RIS illumination. For $N_p=32$, $F=16$, numerical computation yields $\sigma_1^2 = -13.9$ dB and $\sum_{i=1}^{N_a} \sigma_i^2=-11.9$ dB. For $N_p=128$, $F=80$, $\sigma_1^2 = -21.1$ dB for end feed and $\sigma_1^2 = -20.2$ dB for the center feed; we see that the end feed power transfer is 0.9 dB less than the center feed power transfer. The condition number of $\Tm$ for the center feed and the end feed are 5 and 15, respectively. A center feed propagation matrix $\Tm$ is better conditioned than an end feed $\Tm$ due to the inherent geometric symmetry of the center feed.


\begin{figure}[b!]
\centerline{\includegraphics[width=8.25cm]{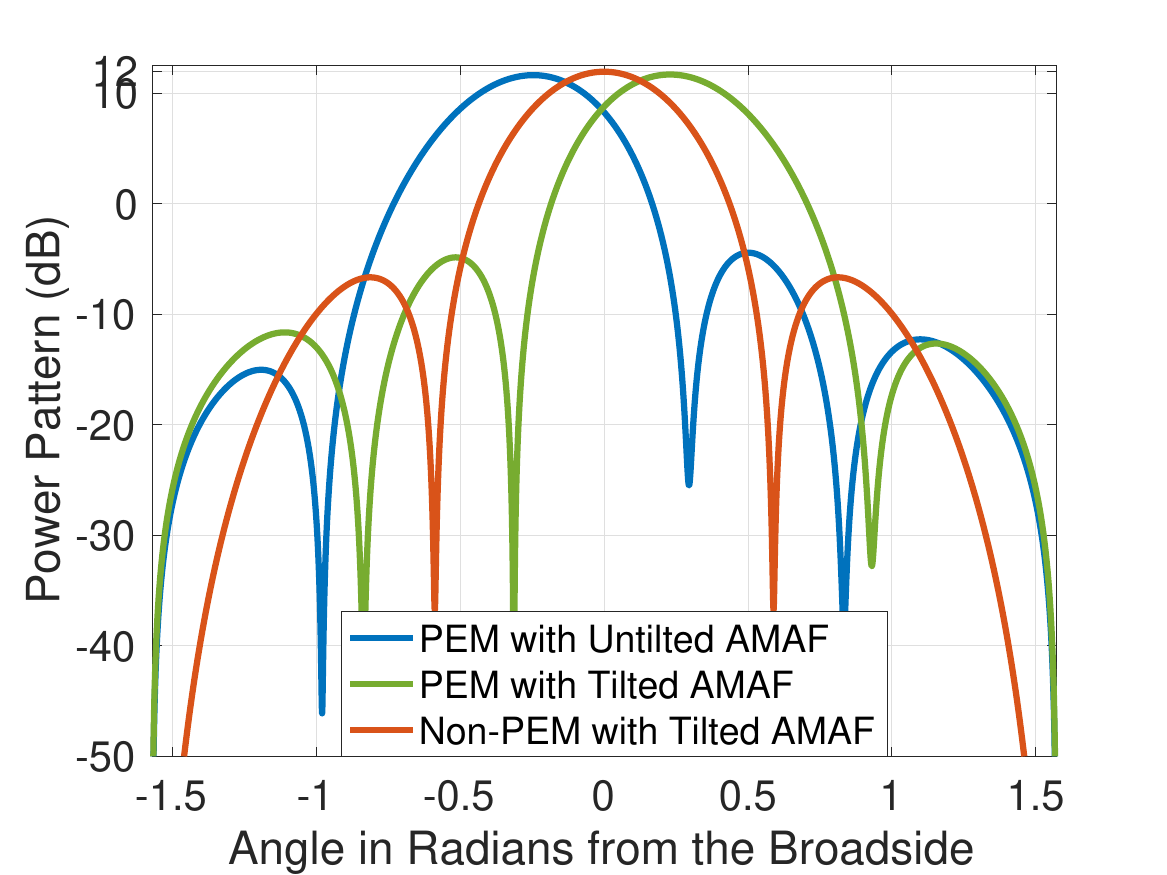}}
	\caption{AMAF Patterns. PEM refers to the principal eigenmode, i.e., when the AMAF is configured along $\vv_1$. Non-PEM refers to AMAF configuration along $\big|\vv_1\big|$. }
	\label{fig:AMAF_pat}
\end{figure}


In Fig. \ref{fig:AMAF_pat}, we plot end feed AMAF radiation patterns. When the AMAF is not tilted toward the RIS center, the principal eigenmode AMAF beam naturally points away from the AMAF boresight towards the RIS center for better illuminating the RIS. When the AMAF is tilted such that its mechanical boresight is pointing to the RIS center (as is conventionally done with feed horn), the principal eigenmode AMAF beam naturally steers on the other side of the AMAF mechanical boresight. This is an interesting natural phenomenon of principal eigenmode (PEM) beamforming, where the tilted AMAF principal eigenmode beam directs more energy to nearby RIS elements for \emph{the maximum AMAF-RIS power transfer}, where not only the distance dependent propagation losses are smaller, but the RIS element factor is also larger due to closer to normal incidences at the RIS elements. This is in contrast to the classical tilted feed horn, where the radiation pattern of the horn antenna remains the same before and after tilting. 

However, the mechanical tilt does not compensate for the fundamental asymmetry of the end feed AMAF-RIS geometry in either case. Therefore, we electronically steer the tilted AMAF beam so that its peak is aligned along its mechanical boresight toward the RIS center, by configuring the AMAF with $\bv=\big|\vv_1\big|$ instead of $\vv_1$, where $\big|\vv_1\big|$ denotes a vector containing the magnitudes of the elements of $\vv_1$. Note that $\big|\vv_1\big|$ is a non-eigenmode AMAF configuration, but it still retains the amplitude characteristics of $\vv_1$, therefore we refer to $\big|\vv_1\big|$ as ``non-PEM''. For the non-PEM pattern in Fig. \ref{fig:AMAF_pat}, the array factor and the element factor contribute 5.93 dB and 6.02 dBi, respectively, resulting into the non-PEM beam peak of 11.95 dBi, which is about 0.4 dB higher than the PEM beam peak due to the larger AMAF element factor at its mechanical boresight. The AMAF array gain is less than the ideal 6.02 dB because the unit norm $\big|\vv_1\big|$ has a taper where the inner two element powers are about 2.6 dB higher than the outer two elements. 

\begin{figure}[ht!]
\centerline{\includegraphics[width=8.25cm]{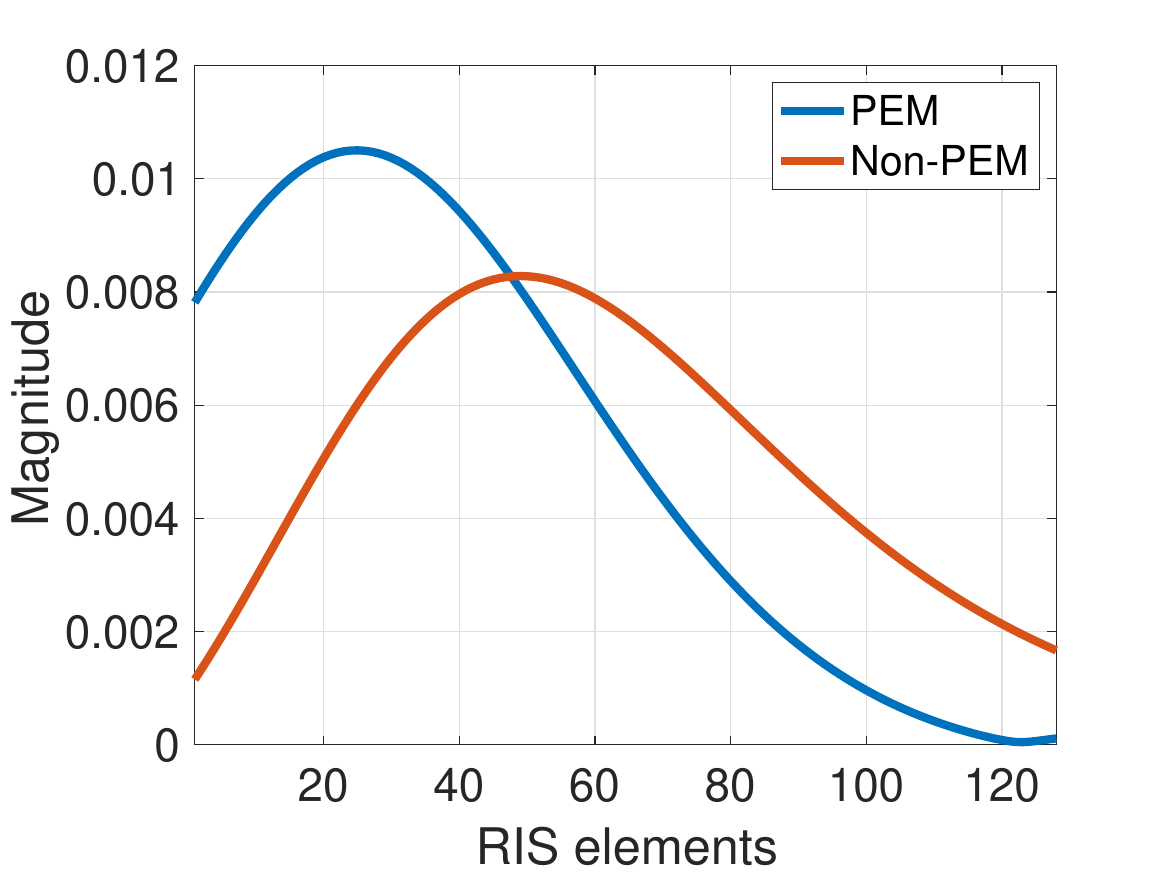}}
	\caption{Magnitude profiles across RIS at $F=110$. PEM and Non-PEM refer to AMAF configurations $\vv_1$ and $\big|\vv_1\big|$, respectively.}
	\label{fig:RIS_mag}
\end{figure}

\begin{figure}[ht!]
\centerline{\includegraphics[width=8.25cm]{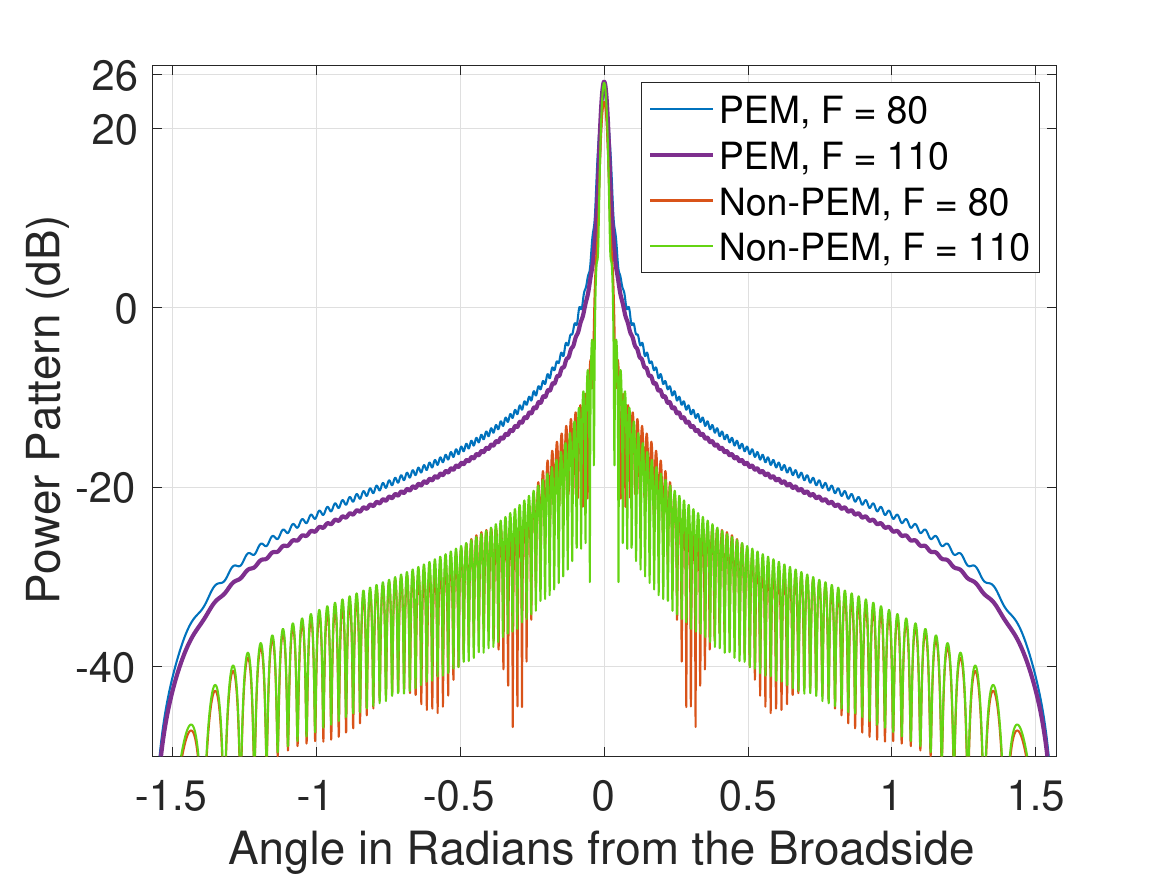}}
	\caption{RIS radiation patterns, tilted end feed AMAF.}
	\label{fig:RIS_pat}
\end{figure}

In Fig. \ref{fig:RIS_mag} we plot the RIS amplitude profile for $\vv_1$ and $\big|\vv_1\big|$ at the focal length $F_{\rm end}=110$ because of its favorable RIS radiation pattern, as we'll see in the next paragraph and in Fig. \ref{fig:RIS_pat}. In Fig. \ref{fig:RIS_mag} we see that the RIS excitation becomes more even with $\big|\vv_1\big|$ compared to $\vv_1$. Notice that even though the tilted AMAF non-PEM beam is pointed to the RIS center, the RIS elements numbered from 42 to 52 (closer to the tilted AMAF) in Fig. \ref{fig:RIS_mag} capture more RF energy than the RIS center elements numbered from 62 to 72, because of smaller distance dependent losses and larger RIS element response near the boresight. However, since $\big|\vv_1\big|$ is the non-PEM, $\big|\big|\Tm|\vv_1|\big|\big|=-23.9 ~{\rm dB}$ compared to the PEM $\big|\big|\Tm\vv_1\big|\big|=\sigma_1^2=-22.6 ~{\rm dB}$, where $\big|\big|.\big|\big|$ is the Euclidean norm of a vector.

We choose the reconfigurable phase shifts at the passive RIS, modeled by a diagonal matrix $\Dm \in \CC^{N_p \times N_p}$, given as $\Dm=\text{diag}(e^{j\phi_1}, \twodots, e^{j\phi_{N_p}})$, such that $\Dm \Tm \bv$ has real positive elements, and denote the resulting vector as $\Dm\Tm\bv = |\Tm \bv|$. Note that $|\Tm\bv|=\sigma_1|\uv_1|$ when $\bv=\vv_1$ for the PEM, and $\big|\Tm\bv\big|=\big|\Tm|\vv_1|\big|$ for the non-PEM. We plot the RIS far-field radiation patterns given by $\big|\av^\herm(\vartheta)\Dm\Tm\bv/\sigma_1\big|^2 E(\vartheta)$ in Fig. \ref{fig:RIS_pat}. We see in Fig. \ref{fig:RIS_pat} that $\big|\vv_1\big|$ results in a much improved RIS pattern due to its more even RIS illumination we saw in Fig. \ref{fig:RIS_mag}, and from an empirical optimization it is best at $F_{\rm end}=110$. 



The $F/D$ ratio of 110/128 = 0.86 is closer to the typical $F/D$ ratio for horn fed parabolic reflectors than the center feed $F/D$ ratio of 0.63. At this $F_{\rm end}=110$, $\big|\big|\Tm|\vv_1|\big|\big|^2 = -23.9$ dB, which is a 3.7 dB power loss compared to the center feed where the optimum $F_{\rm center}=80$ with $\sigma_1^2 = -20.2$ dB. Clearly, the \emph{inherent} geometric asymmetry, the non-eigenmode configuration, and the increased AMAF-RIS distance contribute to a lower AMAF-RIS power transfer for the end feed than for the center feed. Notice that no element of the tilted AMAF faces any RIS element at its boresight, which would be a condition for the globally maximum power transfer between the AMAF and the RIS. Also, sidelobe levels less than -50 dB, as reported for the center feed PEM in \cite[Fig. 3]{ICC2023}, cannot be achieved with an end feed at any $F/D$ ratio.

\section{Conclusions}
\label{conc}

We have quantitatively shown that, for array fed array antenna architectures, center feed has the advantage of higher AMAF-RIS power transfer and beamforming gain (in addition to better and agile beamforming capability and lower profile due to the smaller $F/D$ ratio) over an end feed due to its geometric symmetry, and also because the AMAF and the RIS face each other at their boresights where the element factor is maximum. In the regime of $F/D$ ratio less than one, the AMAF-RIS power transfer deviates from the classical inverse square law of propagation. This is due to the inherent taper of the space feed, which increases as the AMAF and the RIS get closer, where the RIS end elements receive much less power from the AMAF, or vice versa. The AMAF-RIS power transfer behavior exhibits a pattern that depends on the $F/D$ ratio rather than a specific RIS size. These findings apply to both reflectarrays and transmitarrays. Finally, it is fascinating to see yet another physical manifestation of singular value decomposition, where the end-feed AMAF principal eigenmode beam is naturally directed, both before and after tilting, such that the AMAF-RIS power transfer is maximum.


\begin{thebibliography}{1}
\bibliographystyle{IEEEtran}

\bibitem{60s}
D. Berry, R. Malech, and W. Kennedy, ``The Reflectarray Antenna," \emph{IEEE Trans. Antennas Propag.}, vol. 11, no. 6, pp. 645-651, Nov. 1963.

\bibitem{pozar97} 
D. M. Pozar, S. D. Targonski, and H. D. Syrigos, ``Design of Millimeter Wave Microstrip Reflectarrays," \emph{IEEE Trans. Antennas Propag.}, vol. 45, no. 2, pp. 287-296, Feb. 1997.

\bibitem{SWaP-C} 
Z. Zhou, et al., ``Hardware-Efficient Hybrid Precoding for Millimeter Wave Systems With Multi-Feed Reflectarrays," \emph{IEEE Access}, vol. 6, pp. 6795-6806, Jan. 2018.

\bibitem{Webinar}
F. Yang, ``Reconfigurable Intelligent Surface (RIS): A Pearl in Surface Electromagnetics,'' AIDRC Seminar Series [Online]. Available: https://www.youtube.com/watch?v=Y8wRBe8PseA 44' 27'' [Accessed: 10 Mar. 2024].

\bibitem{TICRA1}
R.~E.~Hodges, et al., ``A Deployable High-Gain Antenna Bound for Mars: Developing a new folded-panel reflectarray for the first CubeSat mission to Mars," \emph{IEEE Antennas and Propag. Mag.}, vol. 59, no. 2, pp. 39-49, April 2017.

\bibitem{TICRA}
TICRA, https://www.ticra.com/esa-selects-ticra-to-develop-a-deployable-reflectarray-for-cubesat-applications-with-in-orbit-demonstration/ [Accessed: 10 Mar. 2024].

\bibitem{ICC2023}  
K.~K.~Tiwari and G.~Caire, ``RIS-Based Steerable Beamforming Antenna with Near-Field Eigenmode Feeder," \emph{ICC 2023 - IEEE Int.  Conf. on Commun.}, Rome, Italy, May 2023, pp. 1293-1299.

\bibitem{trees}  
H.~L.~V.~Trees, \emph{Optimum Array Processing: Part IV of Detection, Estimation, and Modulation Theory}, New York: Wiley-Interscience, 2002.

\bibitem{Balanis_antenna_theo}
C. A. Balanis, \emph{Antenna Theory: Analysis and Design}, 3rd Edition, NJ, USA, Wiley, 2016.

\bibitem{ref_wiley_2007}  
J.~Huang and J.~A.~Encinar, \emph{Reflectarray Antennas}, IEEE Press Wiley-Interscience, 2007.

\bibitem{Comp}
P.~Nayeri, A. Z. Elsherbeni, and F. Yang, ``Radiation Analysis Approaches for Reflectarray Antennas [Antenna Designer's Notebook]," \emph{IEEE Antennas and Propag. Mag.}, vol. 55, no. 1, pp. 127-134, Feb. 2013.

\bibitem{SLAcentroid}  
A.~Nafe and G.~M.~Rebeiz, ``On The Phase Center Analysis of Linear Phased-Array Antennas," \emph{IEEE Int. Symp. on Antennas and Propag. \& USNC/URSI Nat. Radio Sci. Meeting}, San Diego, CA, USA, 2017, pp. 2023-2024.

\end{thebibliography}
\end{document}